# Robust hardware Trojan detection leveraging dual-domain features and stacked ensemble learning

Sefatun-Noor Puspa[1*], Abyad Enan[1], Reek Majumder[1], M Sabbir Salek[1], Gurcan Comert[2] and Mashrur Chowdhury[1]

**Abstract**

Cyber-physical systems rely on sensors, communication, and computing, all powered by integrated circuits (ICs). These ICs are vulnerable to malicious hardware attacks, with hardware Trojans being one of the stealthiest threats. Trojans are malicious implants in the circuitry, which are often inserted during design or fabrication stages. This stealthy addition remains dormant until triggered and might cause functional disruptions or sensitive information leakage once triggered. Traditional IC validation methods, such as functional testing and logic analysis, usually fail to capture these subtle anomalies because hardware Trojans are intentionally designed to mimic normal circuit behavior. They often remain dormant under standard test vectors, and the changes they bring into power consumption, timing, and area are minimal. We present a dual-domain feature extraction strategy that combines time-domain features with frequency-domain characteristics of power traces. For the detection, we created an artificial intelligence (AI) based robust Trojan detection framework that integrates traditional machine learning models, such as random forest (RF), gradient boosting (GB), naïve bayes (NB), and deep learning models, such as deep neural network (DNN), long short-term memory (LSTM), and graph neural network (GNN). In this study, we consider these models as baseline AI models to detect Trojan-infected circuits via side-channel power analysis. We employed a stacked ensemble classifier that integrates the distinct strengths of the six baseline models used in this study. After evaluating our stacking ensemble-based detection method on the advanced encryption standard (AES)-Trojan benchmark, which covers diverse Trojan types, the results demonstrate that the ensemble method consistently outperformed all six baseline models. The ensemble-based detection method achieved a macro-averaged area under the receiver operating characteristic (ROC) curve (AUC) of 0.987, while remaining golden-chip-free, meaning it does not rely on a trusted reference IC for baseline comparison. Instead it detects anomalies directly from observable characteristics of untrusted chips, such as side-channel emissions.

**Keywords** Hardware Trojan, Side-channel analysis, Signal processing, Integrated circuits, Hardware security, Artificial intelligence, Stacking ensemble

*Correspondence:
Sefatun-Noor Puspa
spuspa@clemson.edu
[1] Department of Civil Engineering, Clemson University, Clemson, SC 29634, USA
[2] Computational Data Science and Engineering Department, North Carolina A&T State University, Greensboro, NC 27411, USA

## Introduction

Integrated circuits (ICs) form the core of modern cyber-physical systems. The ICs play an important role in applications ranging from automotive control systems and healthcare devices to critical infrastructure and defense systems (Salmani et al. 2013; Tine et al. 2023; Gong et al. 2023). Given their widespread use, ensuring the integrity and security of ICs is necessary.

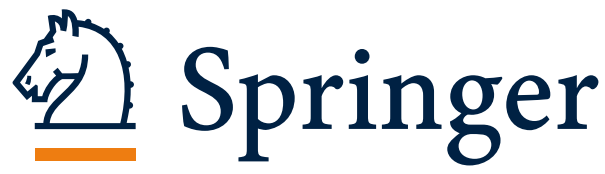

However, the increasing complexity of global supply chains and the sophistication of malicious activity have made ICs vulnerable to hardware-based security threats, including hardware Trojans (Vosatka 2018; Naveenkumar and Sivamangai 2024). Hardware Trojan is a stealthy modification intentionally inserted during the design or manufacturing stages. These hidden Trojans can remain dormant and be triggered with specific inputs, temperature, or time (Vosatka 2018; Naveenkumar and Sivamangai 2024; Lin et al. 2009) (see Fig. 1). Once triggered, a Trojan can cause system failures, unauthorized data manipulation, or other critical security breaches, which makes their timely and reliable detection a crucial area of cybersecurity research (Vosatka 2018; Lin et al. 2009).

Detecting hardware Trojans is complex due to many challenging issues. These malicious modifications often do not alter the functional behavior of the ICs. Moreover, the variations introduced by Trojans in power consumption or area overhead are often minimal and easily masked by design noise and process variability (Hossain et al. 2018; Rad et al. 2010). These characteristics of Trojan attacks make them difficult to detect through conventional testing methods, such as functional testing or logic analysis. In addition, many existing studies adopt a narrow focus, relying solely on individual approaches such as conventional machine learning (Xiang et al. 2019; Hasegawa et al. 2017), without evaluating diverse strategies in a unified framework.

This study presents a unified artificial intelligence (AI) based framework that evaluates various models under identical experimental conditions to address the challenges of hardware Trojan detection. We explored and compared six supervised learning models, which include traditional machine learning classifiers such as random forest (RF), gradient boosting (GB), and naïve bayes (NB); deep learning models including deep neural network (DNN), long short-term memory (LSTM), and graph-based approaches leveraging graph neural network (GNN). All models are trained and tested using a dual-domain feature extraction pipeline that combines time-domain features with frequency-domain spectral characteristics extracted from side-channel power traces. This combination of features together enriches the signal representation and allows the models to detect subtle behavioral changes introduced by Trojan activity.

Beyond the machine and deep learning models, this work introduces a stacked ensemble strategy. We combined all six supervised learning models mentioned above and utilized the outputs of individual classifiers through a meta-learner. This process enhanced the robustness and generalization across diverse Trojan types. The framework is evaluated on the advanced encryption standard (AES)-Trojan benchmark, encompassing multiple

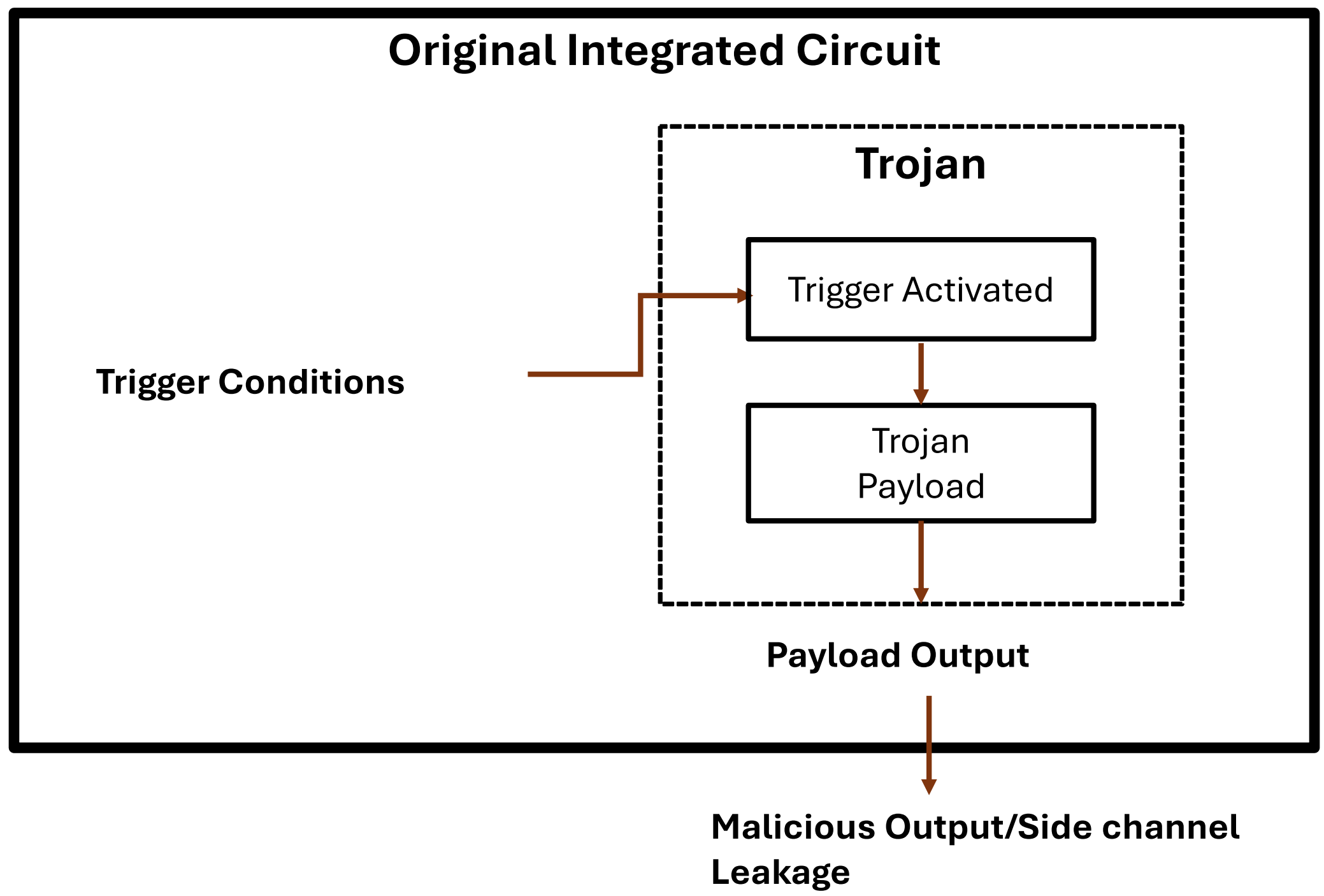

**Fig. 1** Overview of a hardware Trojan's operation mechanism, illustrating trigger activation and malicious payload propagation within an integrated circuit (IC)

Trojan variants with varying activation, complexity and payload type. Our findings demonstrate that combining time-domain and frequency-domain features along with ensemble stacking achieves superior accuracy compared to individual models.

Our key contributions are that we present a unified evaluation framework where six classifiers, traditional, deep, and graph-based, are trained and tested on the same time and frequency-domain features derived for the AES-Trojan benchmarks. Combining these features with a stacking ensemble outperforms every individual baseline model considered in this study.

The remainder of this paper is organized as follows: Sect. "Background and Related Work" offers a detailed review of related work and identifies existing research gaps; Sect. "Framework for Hardware Trojan Detection" introduces our developed multi-model framework and methodology; Sect. "Evaluation" summarizes the experimental setup; Sect. "Results and Discussions" offers a comprehensive discussion of the results of key findings; Sect. "Conclusions and Future Work" outlines future research and concludes by summarizing our contributions and key insights for enhancing IC security.

## Background and related work

Hardware Trojan refers to malicious alterations embedded within the circuitry of ICs. These are often engineered to remain dormant until triggered by specific conditions, such as a predefined input sequence or environmental change. Upon activation, Trojans can severely compromise the circuit's functionality by leaking sensitive information, corrupting data, or initiating denial-of-service (DoS) attacks. Due to their stealthy nature, these attacks introduce negligible power or area overhead, making them particularly difficult to detect during conventional testing procedures (Vosatka 2018; Lin et al. 2009). This section discusses the mechanisms of Trojan insertion and presents reviews of various detection methods explored in existing literature.

### Hardware Trojan insertion

Hardware Trojans can be inserted at various stages of the IC lifecycle. The IC lifecycle can be broadly categorized into two phases: the prefabrication phase and the fabrication phase. The prefabrication phase involves the intellectual property (IP) vendor and IC design house, while the fabrication phase includes the foundry (see Fig. 2). Designers often incorporate third-party IP cores into their IC designs, which may contain hidden Trojans (Naveenkumar and Sivamangai 2024; Ibrahim et al. 2024). Even if the design is initially Trojan-free, poor layout optimization can leave unused spaces in the IC, increasing vulnerability to malicious insertions during fabrication. These insertions, particularly at offshore foundries, are often beyond the control of the original designers and present a serious security risk (Tine et al. 2023; Gong et al. 2023).

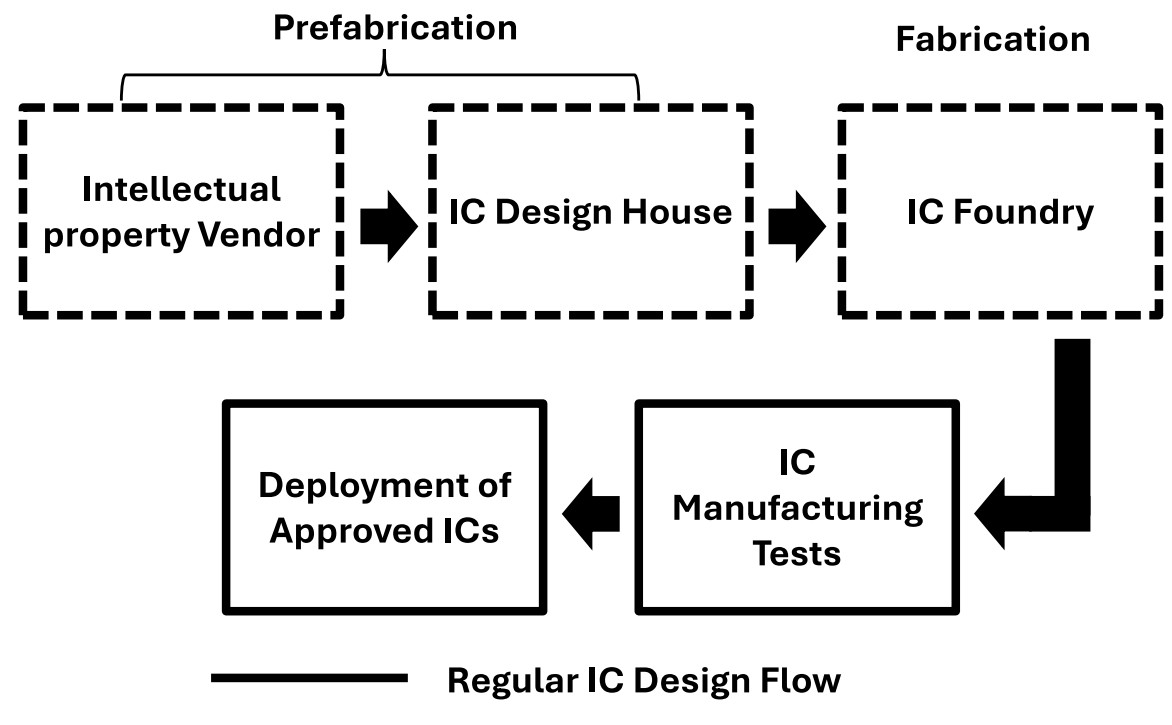

**Fig. 2** IC lifecycle showing prefabrication (design and IP integration) and fabrication (foundry) stages where Trojans can be inserted

### Non-invasive detection techniques

Hardware Trojans are activated by various mechanisms, such as specific input sequences, changes in temperature or predefined time. Depending on their activation conditions, different Trojans exhibit varying sensitivity to different detection techniques. Some Trojans may compromise the IC's performance by slowing down clock frequencies, while others are designed to initiate cyberattacks like DoS or leak confidential encryption keys through covert channels (Lin et al. 2009; Meng et al. 2021; Jin and Makris 2008). This diversity in activation methods and payloads creates a significant challenge for researchers, as no single detection approach is universally effective.

Detecting Trojans before fabrication is essential because post-manufacturing removal of embedded Trojans without destroying the IC is virtually impossible. Once detected at the prefabrication stage, corrective measures, such as redesign or substitution, can be implemented, saving both time and resources (Ibrahim et al. 2024). Existing detection methods can be divided into invasive and non-invasive techniques. Invasive techniques such as reverse engineering and physical inspection involve direct interaction with the circuit, potentially damaging it beyond repair (Vosatka 2018). These methods are often limited to post-fabrication analysis, making them unsuitable for prefabrication detection needs. Non-invasive detection methods do not physically alter the circuit, focusing on observable behaviors (Vosatka 2018). Our research emphasizes non-invasive detection, utilizing side-channel power analysis to identify hidden Trojans based on deviation in IC power consumption. To this end, we review key non-invasive techniques explored in prior research in this section.

1. Functional Testing: This method applies input vectors to the IC and monitors output behavior. Meng et al. (2021) demonstrated that overclocking could detect Trojans by observing bit-flipping at higher clock frequencies. Although functional testing is effective for specific Trojans, it may miss subtle Trojans that do not immediately alter the circuit's functionality (Vosatka 2018).
2. Logic Testing: Logic testing focuses on detecting deviations in the internal logic during Trojan activation. Jin et al. (Jin and Makris 2008) introduced a method using path delay as a metric, comparing delays between Trojan-triggered and non-triggered states. However, this technique may not be effective with Trojans that do not alter the circuit's logical behavior.
3. Formal Verification: This technique evaluates the mathematical correctness of circuit designs. Ibrahim et al. (2024) applied formal verification to detect specific Trojan behaviors by comparing the properties of Trojan-free and Trojan-activated circuits. While accurate, formal verification is computationally expensive and challenging to scale for complex designs.
4. Side-Channel Analysis: To detect Trojans, this technique exploits unintentional emissions, such as power consumption, electromagnetic (EM) emanations, or acoustic signals (Enan and Bhuiyan 2019). For example, Adibelli et al. (2020) used near-field backscattering to detect inactive Trojans by measuring circuit impedance changes. Similarly, Wang et al. (2024) developed an EM sensor array to localize Trojan activity through spatial data collection (Enan and Bhuiyan 2019; Adibelli et al. 2020). However, EM-based methods require specific circuit layouts, limiting their general applicability. Among power-based approaches, hierarchical temporal memory (HTM) has emerged as a brain-inspired method for Trojan detection. Rather than relying on labeled training data, HTM learns streaming input patterns and identifies anomalies through prediction errors. Faezi et al. (2021) applied HTM to the same AES-Trojan dataset used in this study, demonstrating its suitability for real-time detection under streaming power inputs. However, this approach struggles with the detection accuracy, particularly against stealthier or low-leakage Trojans, such as AES-T600 and AES-T1600, with an accuracy of around 60%.

### AI methods in hardware Trojan detection

AI techniques have significantly enhanced hardware Trojan detection due to their ability to effectively capture subtle deviations in IC behaviors through complex data-driven models. Different AI methods have been used to detect hardware Trojans in literature, each demonstrating unique strengths while possessing some inherent challenges.

#### *Traditional machine learning approaches*

*Random Forest (RF):* Xiang et al. (Xiang et al. 2019) utilized a ring oscillator-based side-channel approach combined with an RF classifier to detect hardware Trojans. They have used UART-T800 and UART-T100 Trojan circuits and achieved 100% accuracy in detecting them. Similarly, in a gate-level study, Hasegawa et al. (Hasegawa et al. 2017) applied an RF classifier to structural features extracted from netlists containing Trojans from the Trust-Hub benchmarks, including AES-T100, AES-T300, and AES-T500. They also achieved 100% accuracy. The results of these studies motivated the inclusion of RF as one of our baseline models.

*Gradient Boosting Machines (GBM):* Hayashi et al. (Hayashi and Ruggiero 2025) utilized GB classifiers for Trojan detection in open-source RISC-V designs, analyzing three different Trojan types. While the paper did not explicitly report detection accuracy values, it emphasized GB's superior performance in distinguishing infected designs over other models like decision trees and k-NN.

*Naïve Bayes (NB):* Parmar and Gajjar (Parmar and Gajjar 2023) leveraged Bayesian methods for real-time Trojan detection scenarios, demonstrating computational efficiency but also encountering limitations due to the feature independence assumption. It affected performance when faced with correlated data.

*Support Vector Machines (SVM):* Hu et al. (Hu et al. 2019) applied SVMs effectively for classifying hardware Trojans through side-channel signals, emphasizing their strong generalization capabilities. Although they noted difficulties in parameter tuning and scalability with large datasets.

#### *Neural networks and deep learning techniques*

*Multi-Layer Perceptrons (MLPs):* Pazira et al. (Pazira et al. 2023) explored deep learning techniques, specifically MLPs. They evaluated four Trojan variants from the Trust-Hub RSA benchmark. They have used temperature variation over time (TVOT) and overheated pixel occurrence (OPO) as their features. Their shallow neural network model obtained an average classification accuracy of 82%, which increased to 86% when additional feature sets were incorporated. This exhibits the efficacy of neural networks in identifying thermal deviations introduced by embedded Trojans.

*Convolutional Neural Networks (CNNs):* Roy et al. (Roy et al. 2022) utilized CNNs for side-channel-based Trojan detection, achieving high accuracy through spatial feature extraction of power signals. They evaluated across 12 Trojan types from Trust Hub (AES-T400 to AES-T2000) and achieved an average detection accuracy of 90.48%. Their method outperformed SVM, NB, RF, and HTNet.

*Long Short-Term Memory Networks (LSTMs):* Hoque et al. (Hoque et al. 2024) successfully employed LSTM networks to capture temporal dependencies in power trace data for hardware Trojan detection. The model attained an average accuracy of 98.4%.

*LSTM Autoencoders*: Sumarsono and Masters (Sumarsono and Masters 2023) utilized unsupervised LSTM autoencoders to detect hardware Trojans. They used circuits with always-on and condition-based Trojans. Experiments were conducted on an up-down counter (always-on) and an ethernet router (triggered after 18,000 packets). The model achieved up to 99.87% accuracy. The study focused on the model's strength in detecting delayed and subtle deviations without requiring labeled data.

*Graph Convolutional Networks (GCNs):* Ma et al. (Ma et al. 2025) developed a GNN-based hardware Trojan detection framework that operates at the register-transfer level (RTL). They converted RTL designs into dataflow graphs and extracted multiple categories of features (e.g., node types, structural and control-flow metrics). The authors trained a four-layer GCN model and evaluated it on Trust-Hub benchmarks, including AES-T1000, with a detection accuracy of 99.3%.

*Graph Attention Networks (GATs):* Recent work by Hsiao et al. (Hsiao et al. 2024) explored GATs for identifying structural characteristics through attention mechanisms for Trojan detection. It enables more intricate feature weighting and improved detection accuracy.

#### *Machine learning and deep learning side-channel methods*

Machine learning (ML) assisted side-channel frameworks have also been developed for hardware Trojan detection. Bhatta and Amsaad (Bhatta and Amsaad 2020) applied classical ML models (SVMs, decision trees, neural networks) to the power and EM side-channel signals of hardware Trojan Benchmarks, focusing on AES and RS232 circuits under Trojan-disabled, enabled, and triggered states. Their feature extraction pipeline spanned time, frequency, and wavelet domains, and the classifiers reported detection accuracies exceeding 90% across Trojan instances. Similarly, Nasr et al. (Nasr et al. 2022) introduced a Siamese deep learning architecture for Trojan detection using 60,000 EM traces from the same dataset. They evaluated digital sequential Trojans and showed that a Siamese LSTM achieved 86.78% accuracy, outperforming Siamese GRU (83.59%) and Siamese CNN (73.54%). These works underscore the feasibility of side-channel-based ML/DL detection approaches without reliance on golden chips.

#### *Hybrid and ensemble approaches*

*Hybrid Models:* Xue et al. (Xue et al. 2019) presented clustering ensemble-hardware Trojan detection method (CE-HTDM), a hybrid clustering ensemble-based method for detecting hardware Trojans without relying on trusted golden chips. Their method addresses the challenge of untrustworthy testing, and they combined the results of multiple unsupervised clustering algorithms (e.g., k-means, density-based spatial clustering of applications with noise, hierarchical), outperforming individual techniques. They evaluated on Trust-Hub AES benchmarks and achieved 93.75% detection accuracy.

*Ensemble Learning:* Yoshimi et al. (Yoshimi et al. 2025) presented an ensemble learning based approach for hardware Trojan detection on gate-level netlists. They have used four models: RF, XGBoost, LightGBM, and CatBoost. They evaluated the 32 traditional Trust-Hub benchmarks and 15 additional Trojan-inserted circuits generated via the Trojan insertion tool (TRIT) automatic Trojan insertion framework. Their results showed that CatBoost with optimized borderline synthetic minority oversampling technique (Borderline-SMOTE) achieved the highest average F1-score of 89.31% across 47 benchmarks. The study highlights how ensemble classifiers improve detection performance, even against newly generated and stealthy Trojan variants.

Lavanya and Rajalakshmi (Lavanya and Rajalakshmi 2023) proposed a heterogeneous ensemble learning framework tailored for AES 256 benchmark circuits. Their approach extracted multiple RTL-level behavioral features, including path delay, power consumption, and resource utilization (LUT, flip flops, BRAM, I/O, BUFG), using the Vivado toolchain. To address data imbalance, they applied the synthetic minority oversampling technique (SMOTE) based resampling methods and subsequently performed feature selection via statistical significance testing, cross-correlation, and PCA. Eleven diverse base classifiers spanning Naïve Bayes variants, decision trees, random forest, extra trees, bagging, AdaBoost, and neural networks were integrated through a maximum voting ensemble (MVE) strategy. This heterogeneous design achieved nearly 100% accuracy, F-measure, and AUC across multiple AES benchmark circuits, demonstrating the value of classifier diversity and data augmentation in RTL-level Trojan detection.

Similarly, Rathor et al. (Rathor et al. 2022) developed an ensemble learning model aimed at designing time Trojan vulnerabilities. Their method focused on combining classical machine learning classifiers, support vector machines, random forests, k-nearest neighbors, logistic regression, and XGBoost into a weighted ensemble framework. Evaluated on several Trust-HUB benchmark datasets (e.g., RS232, AES, and other standard Trojan-infested designs), their approach consistently achieved 99 to 100% accuracy and near-perfect F1 scores, including strong cross-dataset validation. By strategically assigning weights to base learners, they enhanced both detection reliability and generalization across different design contexts.

This review highlights the range of AI methodologies explored in recent years for hardware Trojan detection, outlining the unique advantages and challenges of each. However, there remains a critical need for systematic comparative analysis under standardized conditions. To address this, our approach integrates a broad spectrum of AI techniques within a unified evaluation framework. By assessing these models under consistent conditions, we aim to uncover their respective strengths and weaknesses and ultimately develop a robust, adaptable, and effective detection strategy that harnesses the complementary capabilities of each method.

### Perspective on Prior Ensemble Approaches

At the design level, Lavanya and Rajalakshmi (Lavanya and Rajalakshmi 2023) proposed a heterogeneous ensemble framework for AES-256 benchmark circuits. Their approach combined eleven diverse classifiers, including naïve bayes variants, decision trees, random forest, extra trees, bagging, AdaBoost, and neural networks, into a maximum voting ensemble. This strategy, supported by SMOTE resampling and PCA-based feature selection, demonstrated the importance of classifier diversity and advanced preprocessing in RTL-level detection. Similarly, Rathor et al. (Rathor et al. 2022) introduced a weighted ensemble of classical ML classifiers (SVM, RF, KNN, LR, and XGBoost) evaluated on Trust-Hub designs such as AES and RS232. Their work emphasized improving generalization across design-time benchmarks by optimally weighting base learners.

In parallel, side-channel methods have moved detection closer to post-silicon contexts. Bhatta and Amsaad (Bhatta and Amsaad 2020) applied classical ML classifiers, including SVMs, decision trees, and neural networks, to AES and RS232 circuits using power and EM traces. They extracted features across time, frequency, and wavelet domains, demonstrating the feasibility of shallow ML applied directly to side-channel data. More recently, Nasr et al. (Nasr et al. 2022) developed a Siamese deep learning framework for sequential Trojans, training on almost 60,000 EM traces. Their architecture used LSTMs, GRUs, and CNNs to capture temporal similarities between traces, thereby reducing reliance on golden chips.

Our study brings together these two directions in hardware Trojan detection, ensemble-based frameworks at the RTL or design-time level and ML/DL methods applied to side-channel signals. At the RTL level, Lavanya and Rajalakshmi (Lavanya and Rajalakshmi 2023) demonstrated the importance of classifier diversity by integrating eleven models through a maximum voting ensemble supported by resampling and feature selection, while Rathor et al. (Rathor et al. 2022) combined classical classifiers such as SVM, RF, KNN, LR, and XGBoost into a weighted ensemble tailored for design-time benchmarks like AES and RS232. In parallel, side-channel studies have emphasized direct learning from signal traces: Bhatta and Amsaad (Bhatta and Amsaad 2020) applied shallow ML models including SVMs, decision trees, and neural networks to AES and RS232 circuits using time, frequency, and wavelet-domain features, whereas Nasr et al. (Nasr et al. 2022) proposed a Siamese deep learning framework leveraging LSTMs, GRUs, and CNNs to capture temporal dependencies across almost 60,000 EM traces. In contrast, our framework operates at the pre-fabrication stage, where the learning method is developed under the assumption that only power traces are available, without access to RTL-level data. This design allows detection to rely solely on side-channel information, making the approach applicable even when detailed design-level knowledge is absent. It departs from both single-model paradigms and shallow ensembles by introducing dual-domain features that combine statistical measures in the time-domain with spectral metrics in the frequency-domain. We also integrated within a stacked ensemble that unifies shallow learners (RF, GBM, Naïve Bayes) and deep learners (DNN, LSTM, GNN) under a logistic regression meta-classifier. This hybrid design positions our approach as a golden chip–free, side-channel driven strategy that broadens applicability beyond RTL-level abstractions and advances robustness against stealthy Trojans.

These prior studies cannot be benchmarked directly against ours because of differences in abstraction level, datasets, and sensing modalities. RTL-based ensembles use features unavailable in power side-channel analysis, while EM-based Siamese frameworks rely on a different physical measurement. For this reason, we treat them as reference context rather than direct baselines. We revisit this discussion in the results and discussion section, where our experimental findings are positioned relative to these representative approaches.

## Framework for hardware Trojan detection

This section outlines our AI-based methodology for hardware Trojan detection. Building on prior insights into the strengths and limitations of different AI models, we designed an adaptable framework that integrates multiple models, leverages dual-domain feature extraction, and maintains consistency across the data pipeline. To further enhance robustness, we incorporate a stacking ensemble approach that combines the predictive outputs of base classifiers, such as RF, GB, NB, DNN, LSTM, and GNN, into a meta-classifier trained to classify between Trojan-disabled and Trojan-triggered power traces. We chose these supervised models for the ensemble due to their consistent performance. This layered strategy enables the ensemble to exploit the complementary strengths of individual models, improving detection accuracy for both subtle and complex Trojan behaviors. The AI-based Trojan detection approach (see Fig. 3) begins with extracting dual-domain features and combining them with preparing the final dataset for model training. This dataset was then used to train and test six AI models. In addition to individual evaluation, the predictive outputs of the supervised models were stored and later used to train a meta-classifier as part of the stacking ensemble.

### Attack model and assumptions

This work assumes a pre-silicon verification scenario where the designer has access to power traces generated during simulation or emulation of synthesized gate-level netlists. The adversary is presumed to have inserted hardware Trojans during the IC design phase, exploiting third-party IP cores or an untrusted IC designer. These Trojans are stealthy by design, consuming minimal power and area overhead, and remain inactive until triggered.

We assume that the designer does not have physical access to the fabricated chip and must rely solely on observable side-channel signals, particularly dynamic power consumption. The attacker aims to leak secrets (AES keys), cause incorrect outputs, or enable backdoor activation without disrupting the normal functional behavior of the circuit.

The goal of the research is to detect anomalous power patterns linked to Trojan activity using non-invasive analysis. Our detection framework does not require knowledge of the Trojan's internal structure, physical insertion point, or trigger mechanism during inference. Instead, it is trained on labeled power traces, distinguishing between Trojan-disabled and Trojan-triggered conditions. While this approach assumes labeled training data is available, it avoids the need for invasive inspection or golden-chip comparison. This model supports detection at the design validation stage before physical chip fabrication, minimizing risk and cost.

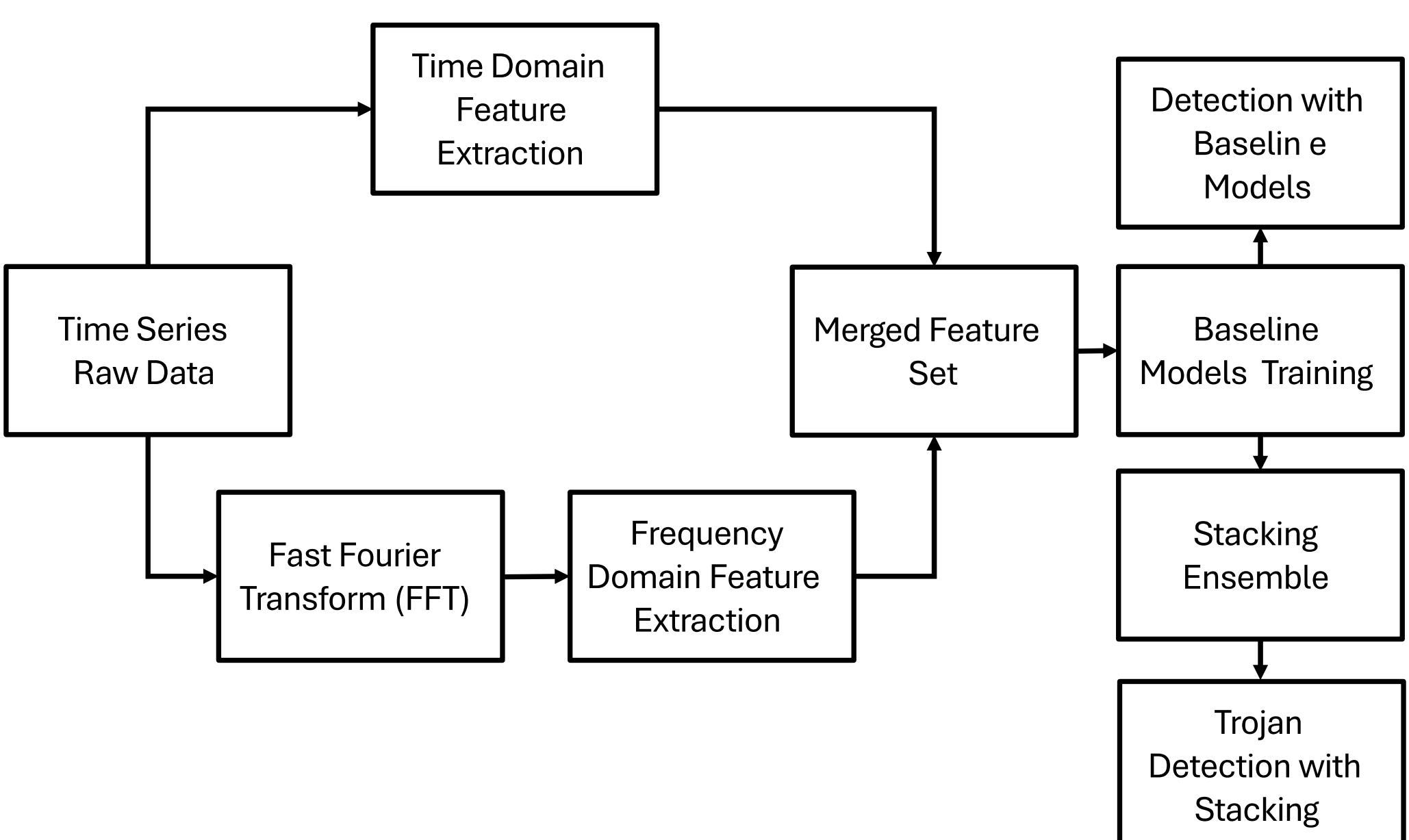

**Fig. 3** Overview of the developed dual-domain (time and frequency) feature extraction and stacked ensemble-based Trojan detection framework

#### Dual-domain feature extraction

To enrich the representation of side-channel signals and improve detection robustness, we employ a feature extraction strategy that combines both time-domain and frequency-domain perspectives. Signal processing plays a critical role in our hardware Trojan detection framework, enabling the extraction of meaningful patterns from complex power consumption data. Time-domain features emphasize short-term deviations due to Trojan activation, while frequency-domain features reveal spectral shifts and harmonic distortions introduced by malicious behavior. This two-step feature extraction process enables training our models to recognize a wide range of Trojan activities. Since power traces are a time series, we first extract features, such as mean, variance, and entropy, from the raw time-domain dataset. Next, fast fourier transform (FFT) is applied to transform the raw power traces into the frequency-domain data, where features such as spectral centroid and total harmonic distortion (THD) help capture deviations caused by Trojan-triggered behaviors.

FFT is a standard algorithm that computes the discrete fourier transform (DFT) more efficiently (Cooley and Tukey 1965). DFT is a mathematical procedure that converts finite, equally spaced sequences to a frequency-domain sequence. However, on a larger scale, DFT becomes very complex. FFT reduced the complexity of DFT. FFT samples a time signal at discrete intervals; after the time-domain signal is broken down into small segments, FFT calculates the DFT of each segment. DFT converts each segment into the frequency-domain. The combined results together create the frequency-domain data from the time-domain. The FFT algorithm we used is the Cooley-Tukey algorithm (Xue et al. 2019). It breaks down a DFT of any composite size into smaller DFT recursively. Each segment uses Eq. 1 to convert a sequence of N numbers from the time-domain to the frequency-domain. Finally, all of them are combined to get the full dataset.

$$X[k] = \sum_{n=0}^{N-1} x[n]e^{-i.2\pi .k.n/N} \quad (1)$$

For all $k = 0, 1, 2, ..N - 1$, where,
$X[k]$: the $k^{th}$ frequency component of the DFT,
$x[n]$: the $n^{th}$ time-domain sample,
$N$: the total number of samples,
$e$: the base of the natural logarithm,
$i$: the imaginary unit,
$2\pi kn/N$: The angle in radians, representing the phase shift for each frequency component.

*Time-Domain Features*: These include statistical metrics such as mean, variance, skewness, kurtosis, entropy, and peak-to-peak amplitude. These features capture short-term fluctuations and signal dispersion in the power traces, which may correspond to transient Trojan activation events. Table 1 describes the features we have extracted from the time-domain dataset. In this table, $n$ represents the total data points, $x$ is the sample, and $x_i$ is the $i^{th}$ data point in the dataset.

*Frequency-Domain Features*: Using the FFT, we extracted frequency-domain features, such as spectral centroid, flatness, harmonic distortion, power bandwidth, and dominant frequency. These features are effective at revealing periodic, low-frequency, and harmonic distortions that may be introduced by stealthy hardware Trojans. Such Trojans often operate subtly by altering switching activity patterns or introducing minor spectral shifts that do not emerge clearly in time-domain analysis. Prior research has established that these frequency-based indicators can expose anomalous behavior hidden in raw power traces, especially for Trojans that rely on rare triggers or timing-dependent activation mechanisms (Iwase et al. 2015; Tang et al. 2020). Table 2 compiled all frequency-domain features extracted after converting the time series dataset to the frequency-domain dataset. In the table, $n$ is the total number of components in the DFT of the signal, M is the number of frequency bands or regions in the spectrum being analyzed, $f_i$ is the $i^{th}$ frequency component, and $X(f_i)$ is the magnitude of DFT of the signal at $i^{th}$ frequency.

Combining time-domain and frequency-domain features creates a richer, more discriminative feature space that enhances model generalization and Trojan detection capabilities. This multi-domain strategy has been shown to improve detection accuracy across various application domains, including vibration fault detection in industrial machinery (Cheng et al. 2018), intrusion detection in industrial control systems (Liu et al. 2024), and biomedical signal analysis in electroencephalogram (EEG) and electrocardiogram (ECG) diagnostics (Shoeb and Guttag 2010). Within hardware security, this dual-domain approach ensures that the models effectively capture and learn transient spikes and persistent spectral behaviors, each reflecting different Trojan behaviors.

#### Multi-model approach

To assess the performance of various detection strategies, we first establish a set of six baseline models, including three traditional machine learning algorithms, RF, GB and NB, and three advanced learning models: DNN, LSTM and GNN. These models were chosen for their complementary strengths. Building on these baseline evaluations, our developed method implements a stacking ensemble classifier that integrates the outputs of the six supervised models, i.e., RF, GB, NB, DNN, LSTM, and GNN. This stacking approach employs a logistic regression meta-classifier to learn from the probabilistic outputs of the base models, improving the robustness and generalization of Trojan detection across diverse activation patterns. Table 3 summarizes the distinct characteristics of each model and their roles within the framework.

**Table 1** Time-domain statistical features extracted from raw power traces

| Features | Definition | Formula |
| --- | --- | --- |
| Mean (μ) | Average power consumption. Detects shifts from normal power usage due to Trojan activation | $\mu = \frac{1}{n}\sum_{i=1}^{n} x_i$ |
| Root Mean Square (RMS) | It measures signal strength and is sensitive to sudden power changes caused by transient Trojans | $\text{RMS} = \sqrt{\frac{1}{n}\sum_{i=1}^{n} x_i^2}$ |
| Variance (σ2) | Spread of power consumption. Highlights irregular power behavior due to Trojan-triggered activities | $\sigma^2 = \frac{1}{n}\sum_{i=1}^{n} (x_i - \mu)^2$ |
| Standard Deviation (σ) | Dispersion of power value. Highlights irregular power behavior due to Trojan-triggered activities | $\sigma = \sqrt{\frac{1}{n}\sum_{i=1}^{n} (x_i - \mu)^2}$ |
| Max | Maximum power value | $x_{\max} = \max(x_i)$ |
| Min | Minimum power value | $x_{\min} = \min(x_i)$ |
| Peak-to-Peak(P2P) | Difference between the maximum and minimum power value. Captures extreme power fluctuations, indicating momentary Trojan activity | $\text{P2P} = x_{\max} - x_{\min}$ |
| Crest Factor | Ratio of peak value to RMS value. Measures signal spikiness, revealing abrupt power increases caused by Trojan triggers | $\text{Crest Factor} = \frac{x_{\max}}{\text{RMS}(x)}$ |
| Skewness | Asymmetry of power distribution. It detects asymmetrical changes in power traces and is useful for finding unexpected behaviors | $\text{Skewness} = \frac{1}{n}\sum_{i=1}^{N} \left(\frac{x_i - \mu}{\sigma}\right)^3$ |
| Kurtosis | It identifies rare spikes in power consumption, suggesting potential Trojan triggers | $\text{Kurtosis} = \frac{1}{n}\sum_{i=1}^{N} \left(\frac{x_i - \mu}{\sigma}\right)^4$ |
| Energy | Total energy consumed | $\text{Energy} = \sum_{i=1}^{N} x_i^2$ |
| Entropy | The randomness of the data. Detects unpredictable behavior in the power signal, a sign of stealthy Trojan activity | $\text{Entropy} = -\sum_{i=1}^{N} p(x_i) \log p(x_i)$ |

**Table 2** Frequency-domain features computed using Cooley-Tukey algorithm

| Features | Definition | Formula |
| --- | --- | --- |
| Spectral Centroid | Center of mass of the power spectrum. Identifies frequency shifts, revealing deviations from normal operation | $\text{Spectral Centroid} = \frac{\sum_{i=1}^{n} f_i \cdot X(f_i)}{\sum_{i=1}^{N} X(f_i)}$ |
| Spectral Bandwidth | The width of the frequency band containing the most power. Detects broad-spectrum changes, indicating abnormal behavior | $\text{Spectral Bandwidth} = \sqrt{\frac{\sum_{i=1}^{n} (f_i - \text{Spectral Centroid})^2 \cdot X(f_i)}{\sum_{i=1}^{n} X(f_i)}}$ |
| Spectral Flatness | Highlights noise-like behavior, signaling potential Trojan activity | $\text{Spectral Flatness} = \frac{\left(\prod_{i=1}^{n} X(f_i)\right)^{\frac{1}{N}}}{\frac{1}{Nn}\sum_{i=1}^{N} X(f_i)}$ |
| Spectral Rolloff | Frequency at which a specific percentage of total spectral energy exists. Identifies shifts in spectral energy distribution, which can be linked to Trojan operations | $\text{Spectral Rolloff} = f_k \quad \text{where} \quad \sum_{i=1}^{k} X(f_i) = 0.85 \sum_{i=1}^{n} X(f_i)$ |
| Spectral Entropy | Entropy of the power spectrum. Measures unpredictability in spectral behavior, indicating stealthy Trojan activity | $\text{Spectral Entropy} = -\sum_{i=1}^{n} P(f_i) \log P(f_i)$ |
| Spectral Contrast | Amplitude differences between spectral peaks and valleys highlight components. Provides insight into circuits about Trojan activation | $\text{Spectral Contrast} = \frac{1}{M}\sum_{m=1}^{M} \left(\max\left(X\left(f_{i_m}\right)\right) - \min\left(X\left(f_{i_m}\right)\right)\right)$ |
| Total Harmonic Distortion | Measure of distortion in the power signal. Detects harmonic interference, suggesting abnormal circuit activity | $\text{THD} = \sqrt{\sum_{k=2}^{K} \left(\frac{X\left(f_{k \cdot f_0}\right)}{X\left(f_{f_0}\right)}\right)^2}$ |
| Harmonic Strength | Magnitudes of the first few harmonics. Measures the strength of harmonics, which can signal periodic Trojan triggers | $\text{Harmonic Strength}_k = X\left(f_{k \cdot f_0}\right)$ |
| Spectral Variability | Variance of the power spectrum. Detects dynamic change in the frequency spectrum | $\text{Spectral Variability} = \frac{1}{N}\sum_{i=1}^{N} \left(X(f_i) - \mu_{X(f_i)}\right)^2$ |

*Random Forest*: RF (Breiman 2001) is an ensemble learning method that constructs several decision trees and combines their predictions to boost overall accuracy. We experimented with model configurations from 8–12 trees and discovered that an ensemble of 10 trees produced the best performance. This setup allowed the model to identify significant patterns in the data without overfitting.

*Gradient Boosting Machines*: GB (Aziz et al. 2020) is an ensemble method that sequentially trains decision trees, with each tree correcting the errors of the previous ones. We experimented with model configurations of 50–150 trees, learning rates of 0.05–0.2, and tree depths of 2–5. We found the optimal configuration for our dataset to be 100 trees, a learning rate of 0.1, and a tree depth of 3. This learning setup was able to fine-tune the predictions at each iteration while also being computationally efficient so that it could be highly responsive to the minute variations in power usage introduced by hardware Trojans.

*Naïve Bayes*: NB is a probabilistic classifier based on Bayes' theorem (Zolnierek and Rubacha 2005) that operates under the presumption of feature independence. i.e., it assumes each feature provides independent information about which class a given input belongs to. We used the Gaussian variant, which models features as normally distributed. Although NB is less dependent on hyperparameter values like other machine learning models, e.g., RF and GB, its probabilistic method helps to perform well when the dataset contains distinct statistical differences between normal and Trojan-affected traces.

*Deep Neural Network*: DNNs (Choi et al. 2020) are highly effective algorithms for modeling complex, nonlinear relationships between features. We used an MLP with one hidden layer, testing neuron counts from 50 to 150 and selecting 100 as optimal. A learning rate of 0.001 and L2 regularization of 0.0001. This also helps avoid overfitting the training data, allowing the model to generalize well to unseen data. The model used ReLU activation and the Adam optimizer, training for up to 200 epochs with early stopping for the prevention of overfitting. Despite its simplicity, the NN effectively captured subtle patterns in both time and frequency-domain features, demonstrating reliable performance across varied Trojan types.

*Long Short-Term Memory*: LSTM networks (Hochreiter and Schmidhuber 1997) are a type of recurrent neural network (RNN) specifically designed to learn from sequential data by preserving long-term temporal dependencies through memory cells and gating mechanisms. In our work, we implemented a bi-directional LSTM architecture to learn both forward and backward dependencies across the power trace sequences. We experimented with hidden dimensions ranging from 64 to 256 and selected a configuration with 128 hidden units and two LSTM layers. The dropout rate was set to 0.5 to prevent overfitting. The model was trained using the Adam optimizer. The initial learning rate was $10^{-4}$ and a weight decay of $10^{-4}$. The overall training was monitored with a scheduler to dynamically reduce the learning rate upon stagnation of validation loss. We used early stopping to avoid overfitting. Although our feature extraction process removed some temporal ordering, the LSTM still proved effective by learning complex relationships between statistical and spectral features. It could identify sequential patterns within the feature sequences that might correspond to Trojan trigger behaviors and activation sequences. Despite the partial loss of fine-grained temporal characteristics during feature aggregation, LSTM was selected for its ability to model residual sequential dependencies among statistical and spectral features.

*Graph Neural Network*: GNN (Ma et al. 2025; Kipf and Welling 2017) allows learning over data structured as graphs, where relationships between instances can be modeled explicitly. In our implementation, each trace was treated as a node in a graph, and edges were formed using both temporal adjacency (sequential traces) and k-nearest neighbor (KNN) similarity based on feature space. The GNN architecture used two graph convolutional layers (GCNConv) with 256 hidden units in the first layer and 2 output units for binary classification. For the training, we used cross-entropy loss and Adam optimizer with a learning rate of 0.001.

**Table 3** Detection models and their unique characteristics

| Models | Strengths in Hardware Trojan Detection |
| --- | --- |
| Random Forest | Effective at detecting faint and noisy patterns in dynamic power data |
| Gradient Boosting | It is very sensitive to even slight changes in power and corrects prediction errors in a sequential manner |
| Naïve Bayes | It assumes feature independence and, thus, works well when there are clear statistical differences between classes in the data |
| Neural Network | Mapping data to high-dimensional regions can identify complex, nonlinear patterns in Trojan-affected power data |
| LSTM | Learns sequential dependencies from time-series power traces, even post-feature aggregation |
| GNN | Exploits structural and relational patterns among traces |
| Stacking Ensemble (Our Method) | Leverages the strengths of multiple models to enhance robustness and accuracy |

*Stacking Ensemble*: To further improve detection accuracy and leverage the complementary strengths of different classifiers, we implemented a stacking ensemble model in our framework. While individual models, such as RF, GB, NB, DNN, LSTM, and GNN, demonstrate unique advantages in capturing distinct aspects of Trojan behavior, their outputs alone may not generalize well across all Trojan types. In our stacking setup (see Fig. 4), we first trained all six supervised models separately using time-domain and frequency-domain features. The class probabilities from these base models were then concatenated to form a new meta-feature set. A logistic regression model served as the meta-classifier, learning to assign optimal weights to each model's output and generate the final decision. This ensemble design is particularly useful when models show conflicting predictions, allowing the meta-learner to balance decisions based on confidence and consistency. Our choice is supported by findings from Hong et al. (Hong et al. 2023), who demonstrated stacking's effectiveness for gate-level Trojan detection. Alternatively, our work extended stacking to side-channel power analysis in the context of non-invasive design stage detection.

## Evaluation

### Dataset description

We used the power side-channel dataset, developed by Faezi et al. (Faezi et al. 2021), to train and evaluate our Trojan detection models. This dataset was specifically designed to support golden chip-free detection and real-time anomaly monitoring. It captures side-channel emissions from an AES-128 encryption core implemented on the SAKURA-G field-programmable gate array (FPGA) board under various Trojan insertion scenarios derived from the Trust-Hub benchmarks (Lin et al. 2009).

The dataset includes 12 distinct Trojan variants (AES-T400 to AES-T2000), each with unique trigger conditions and malicious payloads (key leakage via radio frequency, battery drain, or increased power consumption) (Lin et al. 2009). For each Trojan, signals were recorded under three operational states: Trojan Disabled, Trojan Enabled and Trojan Triggered. Each configuration was tested at seven chip temperatures (25 °C to 85 °C) and two input schemes, fixed input vectors and dynamic "next input = previous output" chaining (Faezi et al. 2021). The dataset comprises over 240,000 labeled traces combining all the circuits, each representing a single AES encryption execution's power signature (Faezi et al. 2021), stored in CSV format. Table 4 presents the dataset used in this study, selected from the publicly available dataset (Faezi et al. 2021) on the Trojan benchmark collection.

### Data preparation pipeline

To ensure consistency across all models, we implemented a standardized data preparation pipeline. It consists of preprocessing, feature extraction, normalization, labeling, and data splitting. Initially, raw power traces were cleaned by removing outliers and smoothing the signals to reduce noise. We then extracted features from both time and frequency domains: time-domain features captured properties like mean, variance, and entropy etc. Frequency-domain features were extracted after converting the raw time-domain signal data to the frequency-domain signal data using fast fourier transform (FFT). Each trace was subsequently labeled as either Trojan-disabled or Trojan-triggered based on its operational condition. Finally, the dataset was split into training (70%), validation (15%), and test (15%) subsets, with no overlap, to ensure unbiased evaluation.

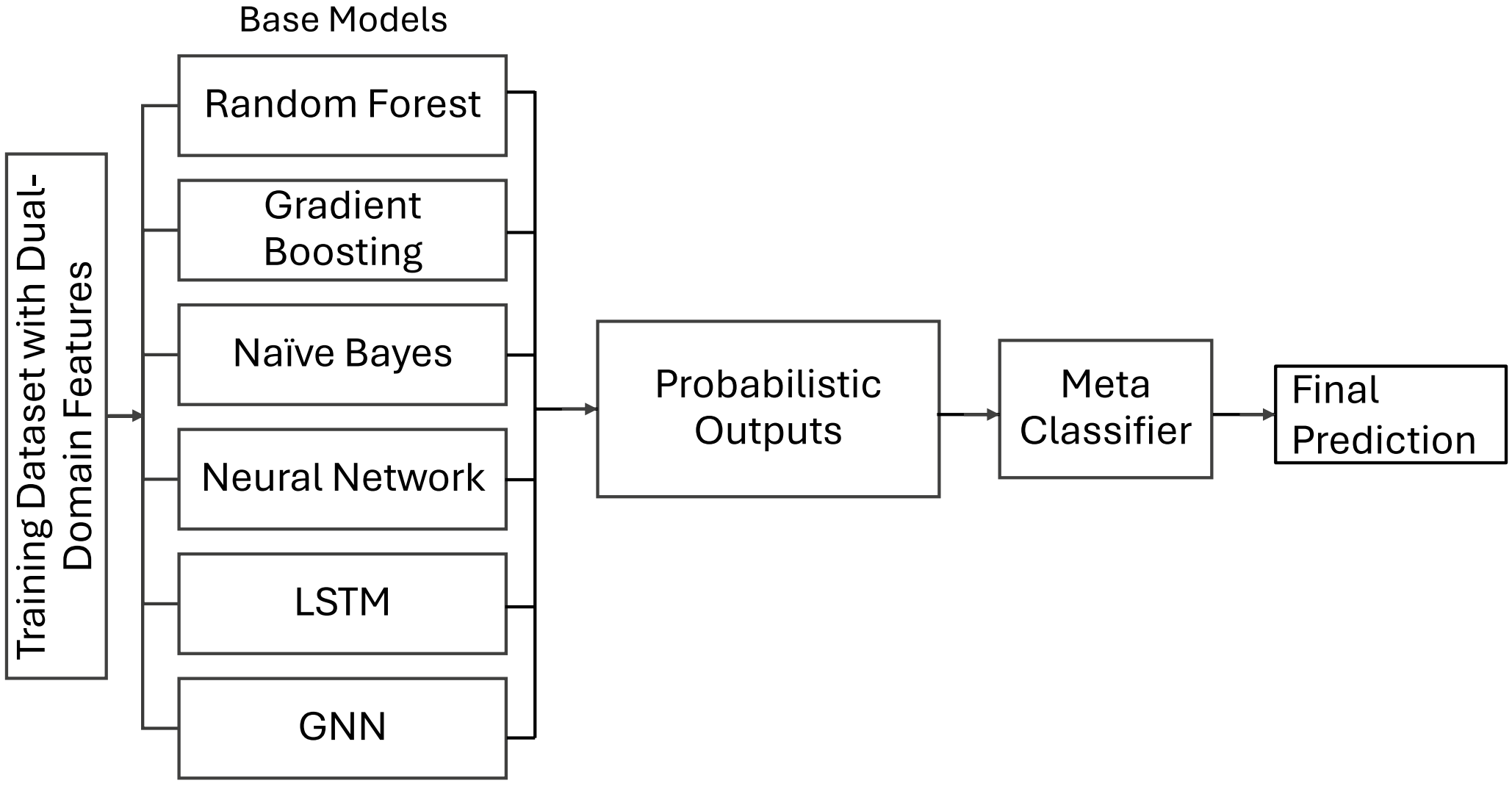

**Fig. 4** Architecture of the stacking ensemble combining six base models (RF, GB, NB, DNN, LSTM, GNN) with a logistic regression meta-classifier

**Table 4** Overview of the Trojan dataset used in this study

| | |
|---|---|
| Source | The dataset is publicly available on the IEEE Dataport |
| Format | The dataset is provided in CSV format. Each file contains time series data on power consumption measurements |
| Physical Parameter | The dataset includes measurements of power consumption |
| Hardware Trojan Benchmarks | The data covers 6 hardware Trojan benchmarks, each representing a unique trigger condition and payload |
| Input Pattern | Data is collected using fixed input values and the "next input = current output" method |
| Chip External Temperature | For this study, we used data collected at 25 °C |

## Design mechanisms and power trace features of selected Trojans

The original dataset (Faezi et al. 2021) contains twelve distinct Trojan variants. For this study, we selected six representative Trojans, T500, T600, T700, T800, T1300, and T1600, based on their diversity in trigger conditions, complexity, and side-channel power leakage characteristics. These six cover a wide spectrum of Trojan activity, ranging from easily distinguishable signatures to low-leakage, stealthy behaviors. This selection was intentional to ensure that the models could be trained on varied signal patterns and later applied to detect similarly structured Trojans in practical scenarios. We included detailed descriptions of these Trojans for a better understanding of their activation mechanisms and feature characteristics. Understanding the Trojans is essential for interpreting model performance and results discussed in Sect. "Results and Discussions".

**AES-T500:** The T500 Trojan uses a shift register that rotates values through a chain of flip-flops. This behavior causes noticeable periodic changes in power due to multiple sequential bit shifts during activation. As a result, extracted time-domain features like variance, standard deviation, peak-to-peak, and crest factor show excellent dispersion between Trojan-triggered and Trojan-disabled states (see Fig. 5).

**AES-T600:** T600 leaks the AES-128 key by creating a direct power-to-ground connection. This introduces low-magnitude power deviations, which are extremely subtle. The resulting side-channel traces are nearly indistinguishable from normal operations in most domains (Naveenkumar and Sivamangai 2024). Time-domain features such as RMS, variance, and entropy, and FFT-based spectral features show overlapping distributions between disabled and triggered traces (see Fig. 6). This makes detection considerably difficult, particularly for simpler classifiers (Lin et al. 2009).

**AES-T700:** T700 employs a code-division multiple access (CDMA) technique to covertly transmit secret data (Lin et al. 2009). The Trojan spreads its signature over time by modulating outputs across multiple clock cycles using pseudo-random codes. This leads to strong harmonic patterns in the frequency-domain, with well-defined peaks. Spectral flatness, dominant frequency, and harmonic distortion serve as reliable indicators.

**AES-T800:** Like T700, T800 uses CDMA encoding but incorporates pseudo-random triggering for added stealth. Despite the additional randomness, it still produces unique spectral behaviors, notably in spectral centroid and power bandwidth (Lin et al. 2009).

**AES-T1300:** This Trojan manipulates intermediate states within the AES-128 key schedule. It injects additional AND gates between input and key bits, subtly altering the logical behavior and causing asymmetric switching activity (Lin et al. 2009). This complexity generates distinguishable distributional shifts in higher-order time-domain features, especially skewness, kurtosis, and entropy (see Fig. 7).

**AES-T1600:** The T1600 Trojan creates unnatural dependencies between key and input bits through a pairwise AND-XOR structure, generating subtle and distributed leakage throughout the AES round computations (Lin et al. 2009). Unlike T1300, its impact is more overlapped, leading to mild RMS and spectral bandwidth increases.

## Results and discussions

This section presents the evaluation outcomes of multiple machine learning and deep learning models applied to hardware Trojan detection. Our experiment spans six Trojan circuits from the AES benchmark (Lin et al. 2009): T500, T600, T700, T800, T1300, and T1600. Each model was evaluated using the same dual-domain feature set and consistent data pipeline to ensure fair comparison, along with a meta-level stacking ensemble that learns from the probabilistic outputs of all six base models. This ensemble approach was designed to aggregate the unique

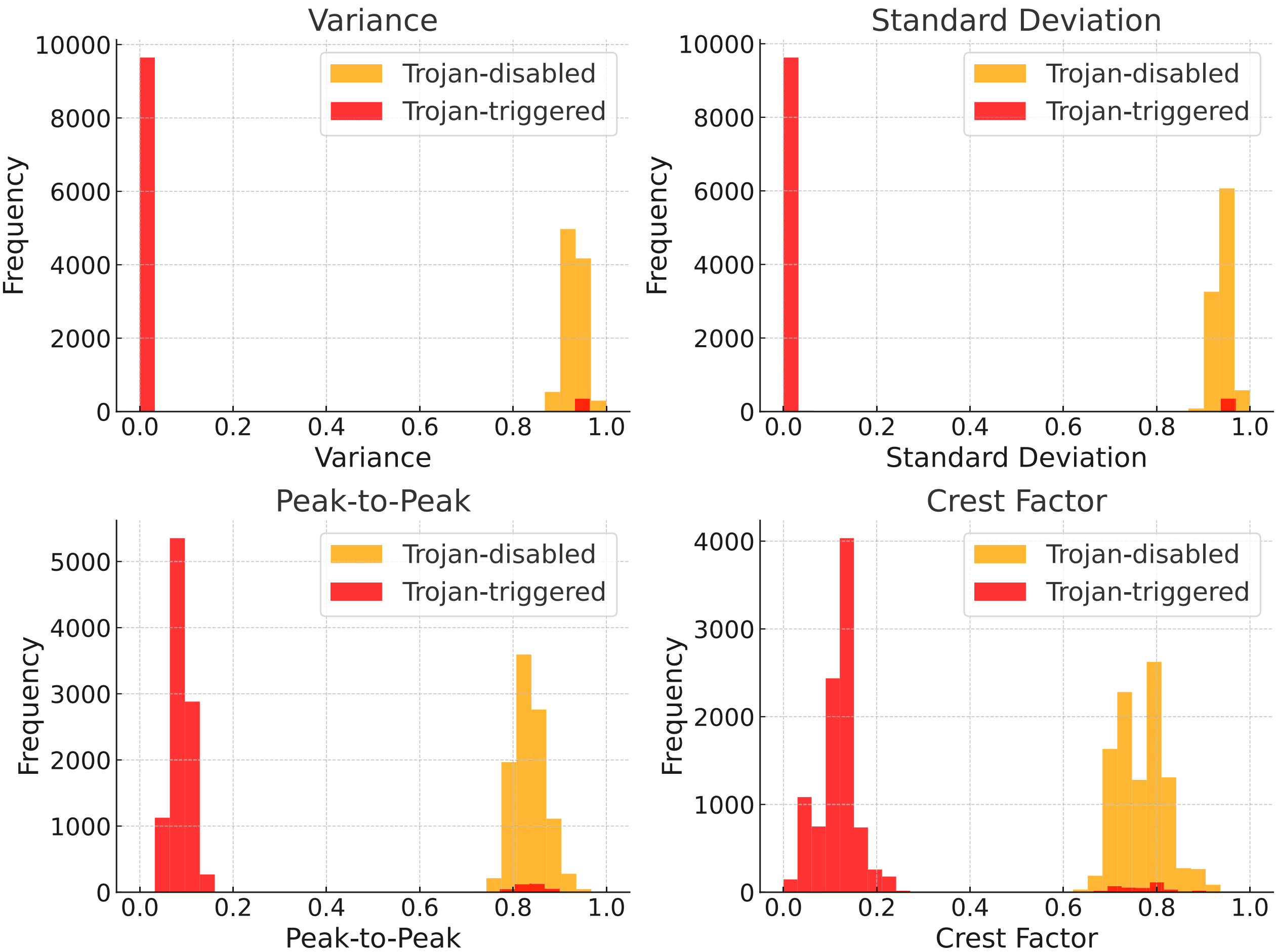

**Fig. 5** Distributions of variance, standard deviation, peak to peak, and crest factor for T500 with the conditions of "Trojan-disabled vs Trojan-triggered" under 25 °C test conditions. Features derived from 10,000 traces

strengths of each baseline classifier and enhance generalization across diverse Trojan behaviors.

## Summary of performance across all models

Table 5 presents a comparison of detection performance across all models and Trojan types. Traditional models such as RF and GB achieve strong detection performance on easily distinguishable Trojans, such as T500, T700, and T800, but struggle on stealthy variants like T600 and T1600. Deep learning models, particularly LSTM and GNN, demonstrate superior adaptability to these low-leakage cases due to their ability to capture temporal and structural patterns. The stacking ensemble, which integrates outputs from multiple base learners, consistently outperforms individual models across nearly all benchmarks, showcasing its robustness and ability to generalize across diverse Trojan behaviors.

### Evaluation of base models

The six supervised classifiers, RF, GB, NB, DNN, LSTM and GNN, serve as baseline models in this study. Table 5 shows the accuracy across all six Trojans with all the models. These models were trained using time and frequency-domain features. Overall, RF and GB performed well on Trojans with strong feature dispersion (T500, T700, T800), with accuracy often exceeding 98%. However, their performance declined on Trojans like T600 and T1600, where signal leakage is low-magnitude and highly overlapping with Trojan-disabled behavior. For example, in T600, RF achieved only 61.1% accuracy, while NB lagged behind at 59.0%. In contrast, for T1300, which exhibits clear feature separation in skewness and kurtosis, all traditional models achieved above 98% accuracy. These results highlight the dependence of traditional classifiers on clear feature separation and their limitations when detecting subtle or covert Trojans.

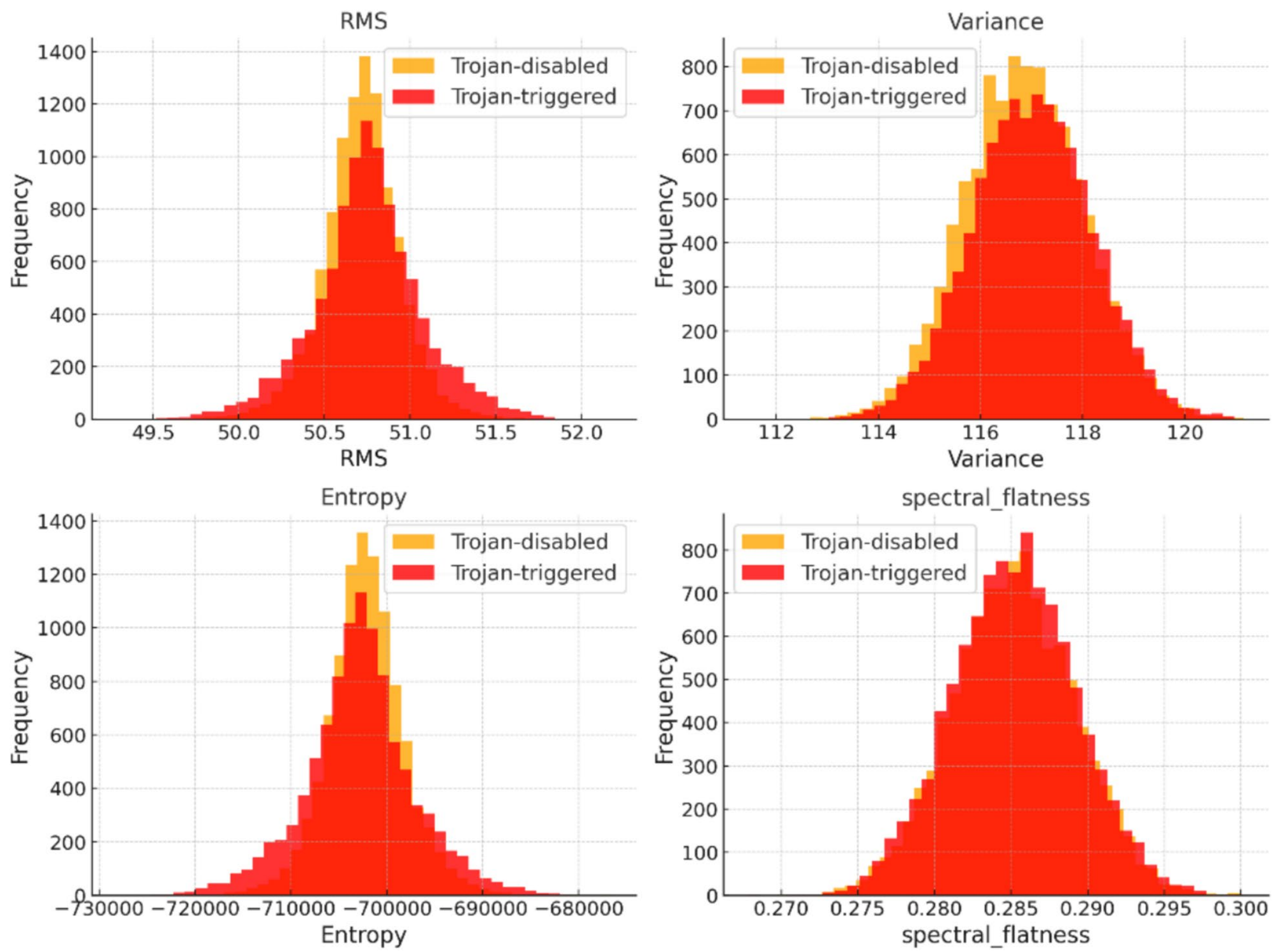

**Fig. 6** Distributions of RMS, variance, entropy, and spectral flatness for T600 with the conditions of "Trojan-disabled vs Trojan-triggered" under 25 °C test conditions. Features derived from 10,000 traces

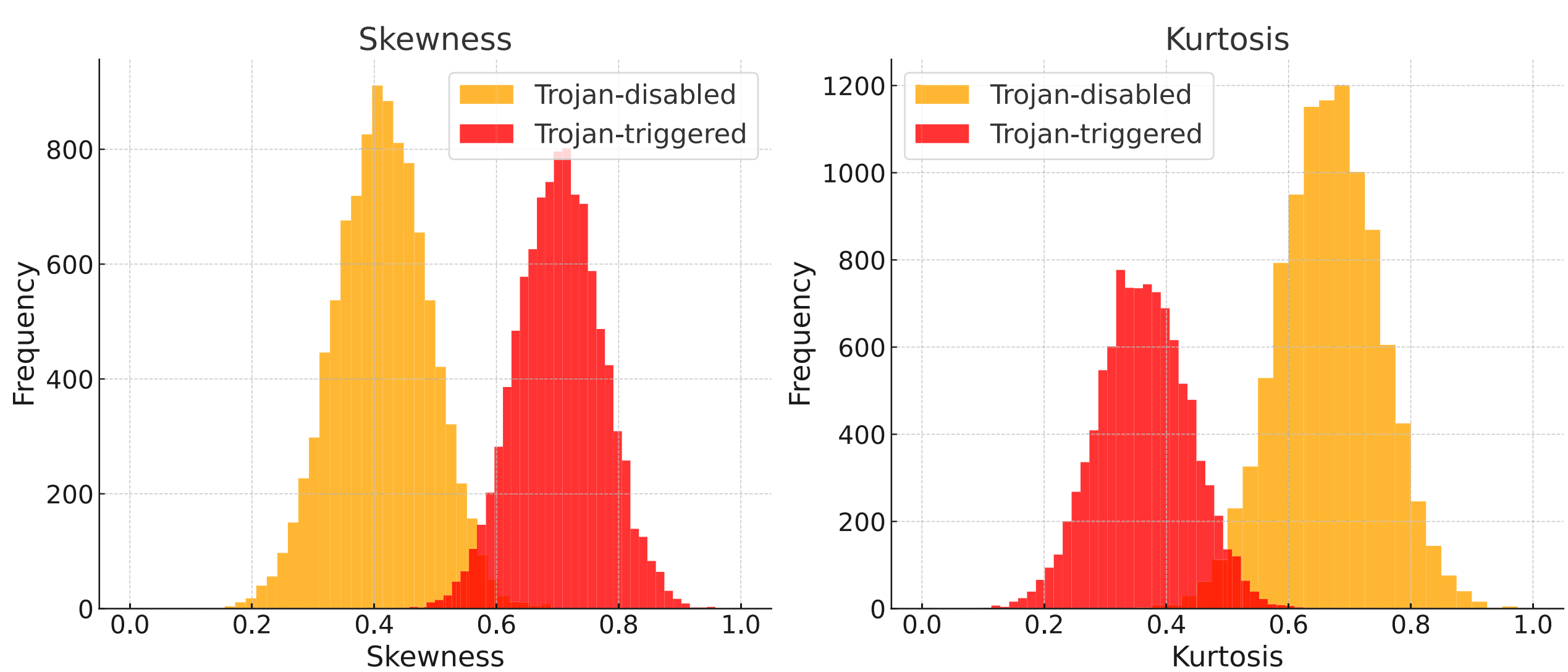

**Fig. 7** Distributions of skewness and kurtosis for T1300 with the conditions of "Trojan-disabled vs Trojan-triggered" under 25 °C test conditions. Features derived from 10,000 traces

**Table 5** Trojan detection performance (%); metrics averaged over 10 runs; test set = 20% of total traces

| Trojan | AI Model | Accuracy (%) ± SD | F1 (%) ± SD | Precision (%) ± SD | Recall (%) ± SD |
|---|---|---|---|---|---|
| T500 | DNN | 98.2 ± 0.2 | 99.9 ± 0.1 | 96.5 ± 0.4 | 98.2 ± 0.2 |
| | GB | 98.2 ± 0.2 | 99.9 ± 0.1 | 96.5 ± 0.4 | 98.2 ± 0.2 |
| | RF | 98.2 ± 0.2 | 99.9 ± 0.1 | 96.6 ± 0.4 | 98.2 ± 0.2 |
| | NB | 98.2 ± 0.2 | 99.9 ± 0.1 | 96.6 ± 0.4 | 98.2 ± 0.2 |
| | LSTM | 98.3 ± 0.2 | 99.8 ± 0.1 | 96.7 ± 0.5 | 98.3 ± 0.2 |
| | GNN | 98.0 ± 0.0 | 99.8 ± 0.0 | 96.2 ± 0.1 | 98.0 ± 0.0 |
| | Ensemble | 99.3 ± 0.7 | 99.7 ± 1.2 | 96.1 ± 3.4 | 97.8 ± 2.1 |
| T600 | DNN | 62.8 ± 0.4 | 64.5 ± 2.0 | 57.5 ± 6.3 | 60.5 ± 2.9 |
| | GB | 56.3 ± 0.8 | 70.4 ± 3.1 | 63.0 ± 0.7 | 63.0 ± 0.7 |
| | RF | 61.1 ± 0.5 | 62.0 ± 0.6 | 57.5 ± 0.8 | 59.7 ± 0.5 |
| | NB | 59.0 ± 0.8 | 65.6 ± 1.2 | 57.3 ± 0.8 | 56.8 ± 0.8 |
| | LSTM | 85.2 ± 1.7 | 82.6 ± 5.3 | 90.3 ± 5.2 | 86.0 ± 1.0 |
| | GNN | 63.0 ± 0.4 | 63.7 ± 0.9 | 60.3 ± 2.2 | 62.0 ± 0.9 |
| | Ensemble | 86.1 ± 3.5 | 79.9 ± 4.1 | 88.2 ± 3.2 | 85.8 ± 4.3 |
| T700 | DNN | 95.7 ± 0.3 | 95.0 ± 1.3 | 96.5 ± 1.8 | 95.7 ± 0.4 |
| | GB | 81.1 ± 0.3 | 91.5 ± 3.0 | 68.8 ± 2.6 | 78.4 ± 0.6 |
| | RF | 90.4 ± 0.3 | 94.4 ± 0.4 | 86.0 ± 0.6 | 90.0 ± 0.4 |
| | NB | 81.8 ± 0.5 | 87.2 ± 0.8 | 74.5 ± 0.6 | 80.3 ± 0.5 |
| | LSTM | 99.8 ± 0.2 | 99.5 ± 0.4 | 99.5 ± 0.4 | 99.6 ± 0.4 |
| | GNN | 92.3 ± 0.8 | 94.5 ± 0.8 | 90.0 ± 0.9 | 92.1 ± 0.8 |
| | Ensemble | 99.5 ± 0.3 | 100.0 ± 0.0 | 99.5 ± 0.4 | 99.5 ± 0.3 |
| T800 | DNN | 95.7 ± 0.3 | 95.0 ± 1.3 | 96.5 ± 1.8 | 95.7 ± 0.4 |
| | GB | 81.1 ± 0.3 | 91.5 ± 0.3 | 68.8 ± 2.6 | 78.4 ± 0.6 |
| | RF | 90.5 ± 0.3 | 94.4 ± 0.4 | 86.0 ± 0.6 | 90.0 ± 0.4 |
| | NB | 81.8 ± 0.5 | 87.2 ± 0.8 | 74.5 ± 0.6 | 80.3 ± 0.5 |
| | LSTM | 99.8 ± 0.2 | 100.0 ± 0.0 | 99.6 ± 0.3 | 99.8 ± 0.2 |
| | GNN | 92.2 ± 0.5 | 94.0 ± 0.8 | 90.1 ± 0.9 | 92.0 ± 0.6 |
| | Ensemble | 99.9 ± 0.1 | 99.4. ± 0.1 | 100.0 ± 0.0 | 99.7 ± 0.5 |
| T1300 | DNN | 98.6 ± 0.2 | 98.4 ± 0.6 | 98.8 ± 0.5 | 98.6 ± 0.2 |
| | GB | 98.3 ± 0.2 | 98.2 ± 0.2 | 98.4 ± 0.3 | 98.3 ± 0.2 |
| | RF | 98.7 ± 0.2 | 98.5 ± 0.3 | 98.8 ± 0.3 | 98.7 ± 0.2 |
| | NB | 98.2 ± 0.2 | 98.0 ± 0.3 | 98.5 ± 0.3 | 98.2 ± 0.2 |
| | LSTM | 99.9 ± 0.0 | 100.0 ± 0.0 | 99.9 ± 0.1 | 99.9 ± 0.0 |
| | GNN | 99.4 ± 0.1 | 99.4 ± 0.1 | 99.4 ± 0.1 | 99.4 ± 0.1 |
| | Ensemble | 99.4 ± 0.6 | 97.6 ± 3.1 | 99.0 ± 2.2 | 98.2 ± 1.8 |
| T1600 | DNN | 63.7 ± 0.6 | 63.2 ± 1.4 | 66.2 ± 6.5 | 64.5 ± 2.6 |
| | GB | 62.1 ± 0.7 | 62.3 ± 1.1 | 61.5 ± 2.6 | 61.5 ± 1.2 |
| | RF | 63.1 ± 0.5 | 62.8 ± 0.6 | 64.3 ± 1.3 | 65.1 ± 0.8 |
| | NB | 61.7 ± 0.6 | 59.0 ± 0.5 | 76.4 ± 0.8 | 66.6 ± 0.5 |
| | LSTM | 91.4 ± 1.0 | 88.5 ± 3.4 | 95.4 ± 3.5 | 91.7 ± 0.9 |
| | GNN | 66.1 ± 0.7 | 65.6 ± 0.8 | 67.8 ± 1.6 | 66.7 ± 0.9 |
| | Ensemble | 90.3 ± 0.5 | 88.2 ± 0.5 | 95.2 ± 5.2 | 76.9 ± 1.8 |

Although our extracted features are largely aggregated, we examined the applicability of sequential learning via LSTM to determine whether it can still capture the hidden temporal correlations. Table 5 shows that the LSTM model achieved high accuracy in detecting complex Trojans, such as T600 and T1600. For T1600, LSTM achieved 91.4% accuracy, outperforming all traditional models. For T500, T700, T800, and T1300, LSTM achieved near-perfect detection accuracy.

The GNN model utilized both temporal and KNN edges to model inter-trace relationships and local data structure. This graph-based architecture was especially beneficial in circuits with dispersed Trojan behavior. For instance, in T1300, GNN achieved 99.4% accuracy, outperforming all other models. Even in the challenging T600 case, GNN achieved 63.0% accuracy, surpassing RF and NB and remaining competitive with LSTM. The GNN's performance on T700 and T800 was also strong, exceeding 92% in both cases.

To further enhance Trojan detection robustness, we integrated a stacking ensemble that leverages the output probabilities from six base classifiers, RF, GB, NB, DNN, LSTM and GNN, and feeds them into a logistic regression meta-classifier. Our Results highlight the superior performance of the stacking approach across all Trojan types. For difficult-to-detect cases, such as T600 and T1600, the ensemble significantly improves overall accuracy, reaching 86.1% and 90.3%, respectively. These results indicate the ensemble's ability to effectively synthesize complementary strengths from different model types, particularly where individual classifiers struggle. Even for well-separated Trojans like T700, T800, and T1300, the stacking model either matches or marginally surpasses the highest-performing individual model, confirming its adaptability across varied Trojan behaviors.

Our experimental results demonstrate how the developed stacked ensemble advances beyond prior RTL-based ensembles and side-channel ML/DL frameworks. While Lavanya and Rajalakshmi's (Lavanya and Rajalakshmi 2023) heterogeneous voting ensemble and Rathor et al.'s (Rathor et al. 2022) weighted design-time ensembles achieved near-perfect performance in RTL settings, their reliance on net-list-level features limits applicability to post-silicon validation. Similarly, Bhatta and Amsaad's (Bhatta and Amsaad 2020) ML classifiers and Nasr et al.'s (Nasr et al. 2022) Siamese deep learning model validated the feasibility of side-channel approaches, but reported accuracy remained below 87% on stealthy Trojans. In contrast, our dual-domain stacked ensemble achieved 90.3% accuracy on the challenging T1600 benchmark, highlighting its superior robustness in low-leakage scenarios. These results underscore the value of combining time- and frequency-domain descriptors with a heterogeneous stack of shallow and deep learners for golden-chip free Trojan detection.

We also include a comparison with Faezi et al.'s HTM (Faezi et al. 2021) detector because it represents a non-invasive approach that is developed for the same AES-Trojan dataset used in our study. The HTM model is trained only on Trojan-free traces and operates on the raw power-waveform-domain reported in their paper. Under those original conditions, HTM attains 0.80 macro-ROC-AUC (Faezi et al. 2021) on the AES-Trojan benchmarks, with perfect scores on high-leakage Trojans (T500, T800) but a sharp drop to ~ 60% on the stealthier Trojans (T600 and T1600) (Faezi et al. 2021). Our study applies a dual-domain feature set and a six-way supervised stacking ensemble. With the benefit of labelled Trojan traces, the stack lifts performance to 0.987 macro-ROC-AUC. Because AUC is threshold-independent, the gain reflects a genuine improvement in classifying Trojan-triggered vs. Trojan-disabled. In scenarios where labels are scarce, HTM remains attractive; when a labelled set is available and feature engineering is feasible, the stacked model offers substantially higher accuracy and reliability.

In addition to accuracy, we also evaluated the end-to-end prediction latency of our stacked ensemble, including all six base models and the logistic regression meta-classifier. For illustration, we report results on the stealthy T600 Trojan, which is one of the most challenging benchmarks due to its low leakage and feature overlap. The average latency was 0.035 ms per trace (35 µs), corresponding to a throughput of nearly 28,845 traces per second. Even in batch mode (batch size = 256), inference required only 8.875 ms per batch, which is orders of magnitude below the 100 ms real-time threshold commonly used in real-time mobility applications (Islam et al. 2018; Karagiannis et al. 2011). These findings demonstrate that, despite involving learning supported by multiple models, the prediction phase is highly efficient. While our method is designed primarily for the pre-fabrication validation stage, where real-time constraints are not critical, the low measured latency indicates feasibility for runtime scenarios.

To assess overall computational efficiency, we also measured the average wall-clock training time across all Trojan types, which was approximately 2.5 minutes. The experiments were conducted on a server equipped with two Intel Xeon Platinum 8358 CPUs (2 × 32 cores, 2.60 GHz), an NVIDIA A100 80 GB GPU, and 250 GB RAM. Figure 8 presents a bar chart summarizing the average training time across all Trojans, showing that the complete ensemble framework remains lightweight and scalable. These findings demonstrate that, despite involving multiple learners, both the training and prediction phases are computationally efficient. While the developed method is designed primarily for the pre-fabrication validation stage, where real-time constraints are less critical, the low measured latency further confirms its feasibility for future runtime and edge-based deployment scenarios.

Overall, the results confirm that the stacked ensemble framework remains both accurate and resource-efficient, making it a promising foundation for future

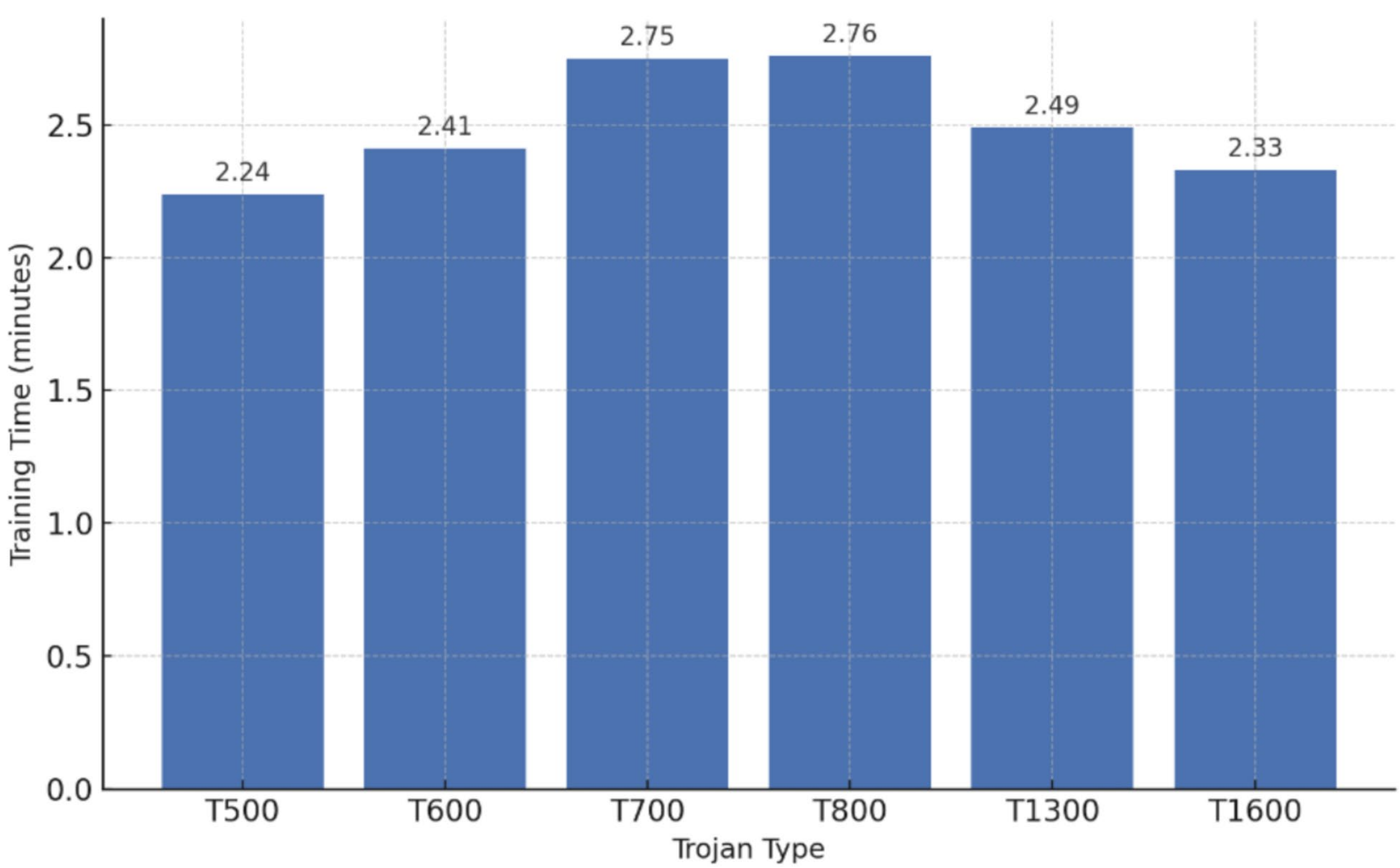

**Fig. 8** Average wall-clock training time of the stacked ensemble model across six Trojan types

extensions in large-scale or edge-deployed Trojan detection systems.

## Limited labeled data evaluation

To evaluate the robustness of the developed framework under limited supervision, we conducted an additional experiment on the stealthy Trojans T600 and T1600, which exhibit the stealthiest behavior among all benchmark circuits we tested. The models were trained using progressively smaller subsets of labeled traces (80%, 50%, 10%, and 1% of the available training data), while maintaining identical testing conditions. As shown in Figs. 9 and 10, the overall accuracy decreased only marginally, by approximately 2-3%, from 80 to 10% training data. Specifically, accuracy for T600 decreased from 85.9% to 81.1%, and for T1600 from 92.9% to 90.1%, demonstrating that the ensemble maintains reliable detection performance even with limited labeled data.

This finding highlights the model's strong generalization capability and supports its potential for golden-chip-free or semi-supervised deployment scenarios where labeled traces are scarce or expensive to obtain.

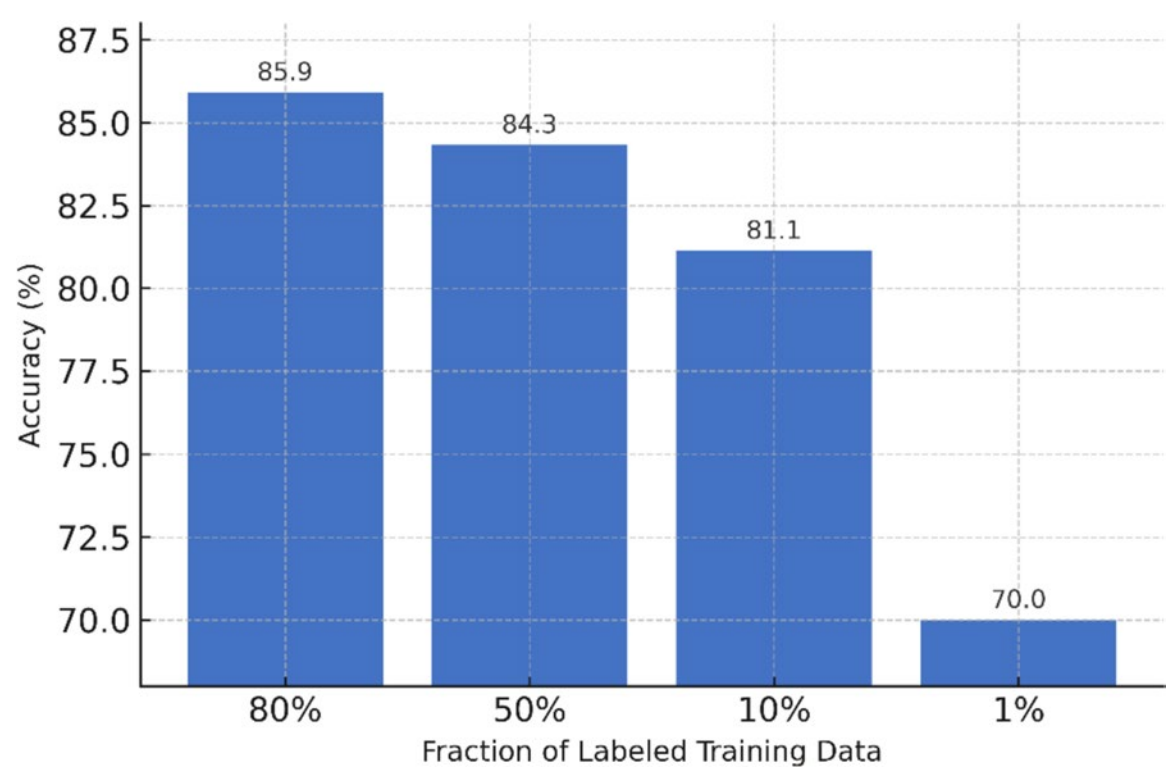

**Fig. 9** Detection accuracy of the developed framework for the T600 Trojan under varying fractions of labeled training data

## Interpreting the results: model behavior and Trojan characteristics

We initially evaluated three traditional machine learning models: RF, GB, and NB. However, due to their limited accuracy in detecting stealthier Trojans, we intentionally introduced a temporal model, such as LSTM and graph-based models like GNN, along with a DNN. Finally, to further enhance detection performance, we employed a stacking ensemble approach that integrates the outputs of multiple supervised classifiers. Our initial use of RF, GB, NB, and DNN provided valuable initial ground. These models performed well with the Trojans, with clear feature dispersion between Trojan-disabled and Trojan-triggered states. However, their accuracy declined significantly for more complex and stealthy Trojans like T600 and T1600. This pattern highlighted the limitations of conventional classifiers when faced with subtle or overlapping features in side-channel power traces.

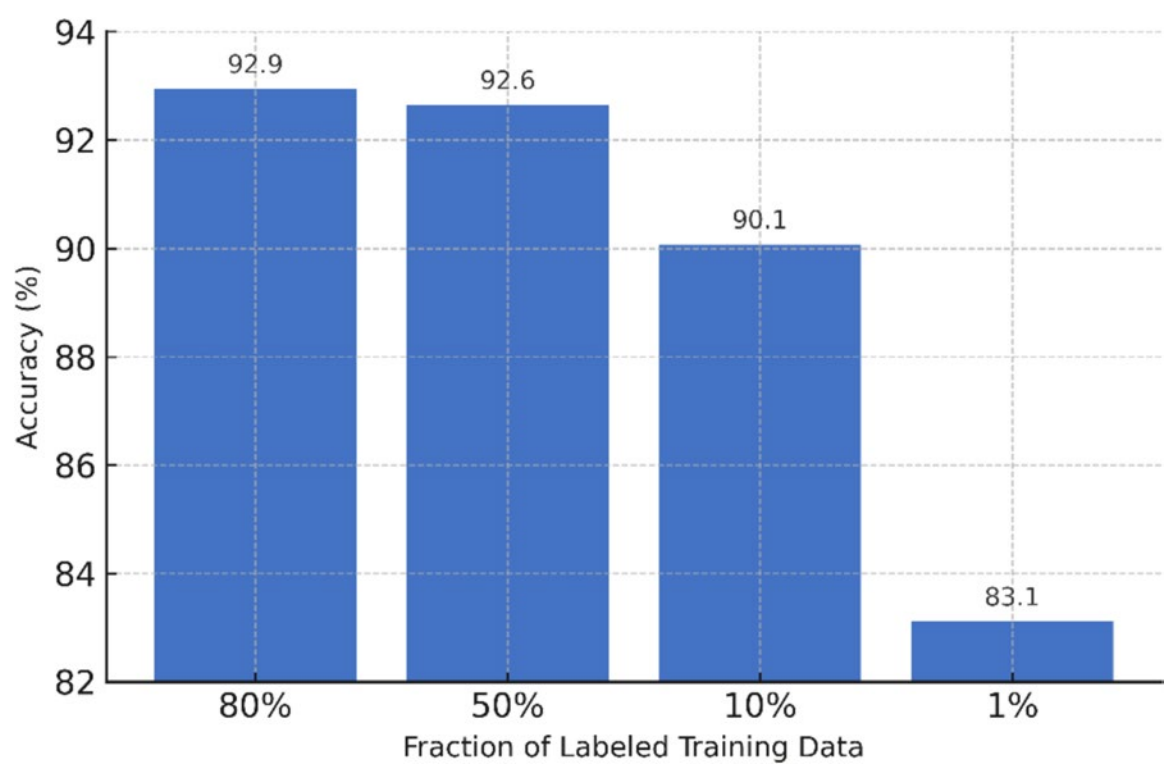

**Fig. 10** Detection accuracy of the developed framework for the T1600 Trojan under varying fractions of labeled training data

To address these limitations, we explored a sequential model. LSTM networks were introduced because of their ability to model time-dependent behavior, even with aggregated features. Even if we lost the raw time-series sequence during preprocessing, the LSTM model could still identify sequential dependencies in the data. This characteristic helped it outperform traditional machine learning models for difficult cases like T600 and T1600.

We introduced a graph-based learning model to further improve detection in complex scenarios. The model learned from individual features, their position, and similarity within the dataset. While the GNN did not achieve high accuracy on challenging Trojans like T600 (63%) and T1600 (66.1%), it still outperformed traditional models like RF and NB. Its strength was more prominent on circuits like T1300, as its features are distributed in a way that creates more noticeable patterns.

Recognizing that no single model consistently performed best across all Trojan types, we integrated a stacking ensemble strategy to synthesize the strengths of diverse classifiers. This approach combined the output probabilities of six base models, including LSTM and GNN, and trained a logistic regression meta-classifier to learn optimal decision boundaries. The stacking ensemble achieved the highest overall accuracy, particularly on previously challenging Trojans, such as T600 and T1600. Its robustness across varying Trojan types demonstrated how combining complementary model behaviors can yield adaptable detection strategies. Together, these findings validate the layered approach of incorporating traditional, sequential, graph-based, and ensemble methods in power-based Trojan detection.

## Conclusions and future work

As integrated circuits become increasingly embedded within critical cyber-physical systems, ensuring hardware-level trustworthiness is essential. This study presents a non-invasive detection framework leveraging side-channel power analysis and signal processing to detect hardware Trojans. By integrating both time-domain and frequency-domain features, our approach enhances the representational capacity of the data and allows AI-driven models to detect even subtle, covert Trojan footprints.

Our results demonstrate that traditional machine learning classifiers such as RF and GB perform well on Trojan circuits with strong statistical separation between classes, such as T500, T700, and T800. However, their effectiveness significantly diminishes in cases with low-amplitude leakage or high feature overlap, as seen in T600 and T1600. Deep learning models, particularly LSTM and GNN, exhibited improved robustness in such scenarios. The GNN achieved over 99% accuracy on T1300 and performed competitively across complex Trojan circuits by leveraging inter-trace dependencies. Most notably, our stacking ensemble framework, which integrates the probabilistic outputs of six diverse models, consistently outperformed individual classifiers. It achieved the highest detection accuracy across nearly all Trojan types, especially improving performance in low-leakage and stealthy scenarios, validating its strength as a generalized and adaptive Trojan detection strategy. In addition to performance consistency across all Trojan families, the framework also demonstrated strong resilience under limited labeled data availability. When the proportion of labeled traces was reduced to as low as 1%, accuracy decreased by only 2-3% for the most challenging Trojans (T600 and T1600), indicating stable learning behavior even with minimal supervision. These methods can be applied to future Trojans that affect dynamic power consumption, even in the most subtle ways. Importantly, our detection framework is applicable at the pre-fabrication stage of the product life cycle and does not require invasive hardware access, enabling more secure and scalable validation processes.

Despite these strengths, several limitations remain that could guide future work. A current limitation of our framework is its dependence on labeled power traces, which may not always be practical in industrial environments. While supervised models deliver high accuracy when required data are available, future work will explore semi-supervised and unsupervised methods to reduce the need for large, labeled datasets while maintaining robust detection.

Another limitation is the lack of support for real-time scenarios. The development of lightweight ensemble variants for resource-constrained or real-time scenarios is required. Our current design is targeted at pre-fabrication validation, and hence time constraints were not of concern. For our future work on cases where real-time detection is required, we plan to investigate cascading ensembles, where lighter models (e.g., RF, GBM, or LSTM) are applied first, and the full stacked ensemble is only invoked when necessary. Additionally, techniques such as pruning, quantization, and knowledge distillation could further reduce computational cost, enabling practical deployment in edge or time-critical settings.

Finally, we acknowledge that manual feature extraction (e.g., FFT and statistical descriptors) may not scale efficiently for ultra-large-scale ICs with multiple power domains and overlapping activity. Future work will therefore investigate automated feature extraction, sparse selection, and hardware-accelerated pipelines to enhance the scalability and portability of the models to help them adapt to more complex SoC environments.

Beyond these immediate extensions, future research directions include the exploration of additional signal measures, such as entropy-based dynamics, stimulus-driven Trojan activation via fault injection or acoustic stimulation, and multi-modal Trojan detection schemes that combine logic testing with side-channel analysis. Beyond detection, supply chain security remains a critical concern. Integrating blockchain technologies to track and verify the integrity of IC design, manufacturing, and delivery could offer a transparent and tamper-resistant foundation for ICs. Smart contracts could automate IP validation and ensure the trust of third-party components, which could also be evaluated in future work.

In summary, this work contributes a modular, adaptable, and AI-enhanced framework for Trojan detection. It can advance the resilience of cyber-physical systems and infrastructure while securing critical embedded systems in the face of evolving hardware-level threats.

**Abbreviations**

| AI | Artificial intelligence |
|---|---|
| IC | Integrated circuit |
| IP | Intellectual property |
| CPS | Cyber-physical systems |
| DoS | Denial of service |
| RTL | Register transfer level |
| EM | Electromagnetic |
| DFT | Discrete fourier transform |
| FFT | Fast fourier transform |
| KNN | K-nearest neighbor |
| CSV | Comma-separated values |
| AUC | Area under curve |
| MLP | Multi-layer perceptron |
| TVOT | Temperature variation over time |
| OPO | Overheated pixel occurrence |
| HTM | Hierarchical temporal memory |
| TRIT | Trojan insertion tool |
| FPGA | Field programmable gate array |
| AES | Advanced encryption standard |
| CE-HTDM | Clustering ensemble hardware Trojan detection method |
| RF | Random forest |
| ROC | Receiver operating characteristic |
| GB / GBM | Gradient boosting / gradient boosting machines |
| NB | Naïve Bayes |
| SVM | Support vector machine |
| DNN | Deep neural network |
| CNN | Convolutional neural network |
| LSTM | Long short-term memory |
| GNN | Graph neural network |
| GCN | Graph convolutional network |
| GAT | Graph attention network |
| RMS | Root mean square |
| THD | Total harmonic distortion |
| P2P | Peak-to-peak |
| ECG | Electrocardiogram |
| EEG | Electroencephalogram |
| RISC-V | Reduced instruction set computer five |
| PCA | Principal component analysis |
| UART | Universal asynchronous receiver–transmitter |
| GRU | Gated recurrent unit |
| LUT | Look-up table |

**Acknowledgements**
This work is based upon the work supported by the National Center for Transportation Cybersecurity and Resiliency (TraCR) Grant No. 2026310 and 2026311 (a U.S. Department of Transportation National University Transportation Center) headquartered at Clemson University, Clemson, South Carolina, USA, and National Science Foundation (NSF) Grant No. 2131080. Any opinions, findings, conclusions, and recommendations expressed in this material are those of the author(s) and do not necessarily reflect the views of TraCR, and the U.S. Government assumes no liability for the contents or use thereof.

**Author contributions**
The authors confirm their contribution to the paper: study conception and design: Sefatun-Noor Puspa, Abyad Enan, Reek Majumdar, M. Sabbir Salek, Gurcan Comert, Mashrur Chowdhury; data collection: Sefatun-Noor Puspa, Abyad Enan; analysis and interpretation of results: Sefatun-Noor Puspa, Abyad Enan, Reek Majumdar, Mashrur Chowdhury; draft manuscript preparation: Sefatun-Noor Puspa, Abyad Enan, Reek Majumdar, M. Sabbir Salek, Gurcan Comert, Mashrur Chowdhury. All authors reviewed the results and approved the final version of the manuscript.

**Funding**
This work is supported by USDOT Grants No. 2026310 and 2026311, and NSF Grant No. 2131080.

**Data availability**
All datasets and models used in this study are publicly available at the following Zenodo repository: https://zenodo.org/records/17519027?token=eyJhbGciOiJIUzUxMiIsImlhdCI6MTc2MjIzMDIzNywiZXhwIjoxNzkzNzUwMzk5fQ.eyJpZCI6IjYxZTM4NThlLWVlYTItNDM4YS1hOTBkLWFkYmFjNWEyNmNmMCIsImRhdGEiOnt9LCJyYW5kb20iOiI0ZTcyMTJhMzJkMmVmYWVkMmYwYzQxMWMyYTdiZWZhOSJ9.ANq3pwlJD49Rcf0v_yxZ-4JS8HF2_k_qpcoTD3HTbfGbR6_MLFflJndMJEhqJTkqI9kT8rN_FpTQsr7iZe5x7w.

## Declarations

**Competing interests**
The authors declare that they have no competing interests.

## Publisher's Note

**Sefatun-Noor Puspa** is a Ph.D. student in the Glenn Department of Civil Engineering, Clemson University, Clemson, SC 29634, USA

**Abyad Enan** is a Ph.D. student in the Glenn Department of Civil Engineering, Clemson University, Clemson, SC 29634, USA.

**Reek Majumdar** is a Director of Engineering at Ipsos MMA in New York. He earned his doctorate in Civil Engineering from Clemson University, Clemson, SC 29634, USA.

**M Sabbir Salek** is a senior engineer at the USDOT National Center for Transportation Cybersecurity and Resiliency (TraCR), headquartered in the Glenn Department of Civil Engineering, Clemson, SC 29634, USA

**Gurcan Comert** is an Associate Professor in the Computational Data Science and Engineering Department, North Carolina A&T State University, Greensboro, NC 27411, USA.

**Mashrur Chowdhury** is the Eugene Douglas Mays Chair of Transportation in the Glenn Department of Civil Engineering, Clemson University, Clemson, SC 29634, USA